\documentclass[twocolumn,showpacs,preprintnumbers,amsmath,amssymb,floatfix,superscriptaddress,PRC]{revtex4
}
\usepackage{graphicx}
\usepackage{dcolumn}
\usepackage{bm}

  \def\nuc#1#2{\relax\ifmmode{}^{#1}{\protect\text{#2}}\else${}^{#1}$#2\fi}
  \def\itnuc#1#2{\setbox\@tempboxa=\hbox{\scriptsize\it #1}
    \def\@tempa{{}^{\box\@tempboxa}\!\protect\text{\it #2}}\relax
    \ifmmode \@tempa \else $\@tempa$\fi}

\begin{document}
\preprint{APS/123-QED}

\title{The $^{14}$C$(n,\gamma)$ cross section between 10 keV and 1 MeV}

\author{R.~Reifarth}
\altaffiliation[present affiliation:]{ GSI, Planckstr. 1, 64291 Darmstadt, Germany.}
\affiliation{Los Alamos National Laboratory, Los Alamos, New Mexico, 87545, USA}
\author{M.~Heil}
\altaffiliation[present affiliation:]{ GSI, Planckstr. 1, 64291 Darmstadt, Germany.}
\affiliation{Forschungszentrum Karlsruhe, Institut f\"ur Kernphysik,
            P.O. Box 3640, D-76021 Karlsruhe, Germany}
\author{C.~Forss\'en}
\affiliation{Lawrence Livermore National Laboratory, Livermore, California, 94550, USA}
\affiliation{Fundamental Physics, Chalmers University of Technology,
  412~96 G\"oteborg, Sweden}
\author{U.~Besserer}
\affiliation{Forschungszentrum Karlsruhe, Tritiumlabor,
             P.O. Box 3640, D-76021 Karlsruhe, Germany}
\author{A.~Couture}
\affiliation{Los Alamos National Laboratory, Los Alamos, New Mexico, 87545, USA}
\author{S.~Dababneh}
\altaffiliation[present affiliation:]{ Faculty of Science, Balqa Applied University, P.O.Box 7051, Salt 19117, Jordan.}
\affiliation{Forschungszentrum Karlsruhe, Institut f\"ur Kernphysik,
            P.O. Box 3640, D-76021 Karlsruhe, Germany}
\author{L.~D\"orr}
\affiliation{Forschungszentrum Karlsruhe, Tritiumlabor,
             P.O. Box 3640, D-76021 Karlsruhe, Germany}
\author{J.~G\"orres}
\affiliation{University of Notre Dame, Physics Department,
             Notre Dame, IN 46556, USA}
\author{R.C.~Haight}
\affiliation{Los Alamos National Laboratory, Los Alamos, New Mexico, 87545, USA}
\author{F.~K\"appeler}
\affiliation{Forschungszentrum Karlsruhe, Institut f\"ur Kernphysik,
            P.O. Box 3640, D-76021 Karlsruhe, Germany}
\author{A.~Mengoni}
\affiliation{CERN, CH-1211 Geneva 23, Switzerland}
\author{S.~O'Brien}
\affiliation{University of Notre Dame, Physics Department,
             Notre Dame, IN 46556, USA}
\author{N.~Patronis}
\altaffiliation[present affiliation:]{ Instituut voor Kern- en Stralingsfysica, K. U. Leuven, Celestijnenlaan 200D, B-3001 Leuven, Belgium.}
\affiliation{Nucl. Phys. Lab., Department of Physics, The University of Ioannina,
            45110 Ioannina, Greece}
\author{R.~Plag}
\altaffiliation[present affiliation:]{ GSI Darmstadt, Planckstr. 1, 64291 Darmstadt, Germany.}
\affiliation{Forschungszentrum Karlsruhe, Institut f\"ur Kernphysik,
            P.O. Box 3640, D-76021 Karlsruhe, Germany}
\author{R.S.~Rundberg}
\affiliation{Los Alamos National Laboratory, Los Alamos, New Mexico, 87545, USA}
\author{M.~Wiescher}
\affiliation{University of Notre Dame, Physics Department,
             Notre Dame, IN 46556, USA}
\author{J.B.~Wilhelmy}
\affiliation{Los Alamos National Laboratory, Los Alamos, New Mexico, 87545, USA}

\date{\today}
\begin{abstract}
The neutron capture cross section of $^{14}$C is of relevance for
several nucleosynthesis scenarios such as
inhomogeneous Big Bang models, neutron induced CNO cycles, and
neutrino driven wind models for the $r$ process. The
$^{14}$C$(n,\gamma)$ reaction is also important for the validation
of the Coulomb dissociation method, where the $(n,\gamma)$ cross
section can be indirectly obtained via the time-reversed process. So far, the
example of $^{14}$C is the only case with neutrons where both, direct measurement
and indirect Coulomb dissociation,
have been applied. Unfortunately, the interpretation is obscured
by discrepancies between several experiments and theory. Therefore,
we report on new direct measurements of the $^{14}$C$(n,\gamma)$
reaction with neutron energies ranging from 20 to 800~keV. 


\end{abstract}

\maketitle

\section{Introduction}
Inhomogeneous big bang models \cite{ApH85} offer the possibility to
bridge the mass gaps at $A=5$ and 8 and to contribute substantially
to the synthesis of heavier nuclei. The suggested reaction sequence
\cite{AHS88,MaF88} for this outbreak is
$^7$Li($n, \gamma$)$^8$Li($\alpha, n$)$^{11}$B($n, \gamma$)$^{12}$B($\beta^-$)$^{12}$C.
Subsequent neutron captures on $^{12}$C and $^{13}$C will then lead
to the production of $^{14}$C, which has a half-life of 5700$\pm$30~yr \cite{AjS86a}. 
On the time scale of big bang nucleosynthesis
$^{14}$C can be considered as stable and further proton, alpha,
deuteron, and neutron capture reactions on $^{14}$C will result in
the production of heavier nuclei with $A \geq 20$ \cite{AHS88}. Due 
to the high neutron abundance the $^{14}$C$(n,\gamma)$$^{15}$C reaction 
is expected to compete strongly with other reaction channels.

The $^{14}$C($n, \gamma$)$^{15}$C reaction plays an
important role in the discussion of neutron induced CNO cycles
\cite{WGS99} during $s$-process nucleosynthesis. Such $s$-process
scenarios are characterized by comparably low neutron densities,
resulting in neutron capture times, which are slow compared to 
typical $\beta$ decay half lives and are associated with the He 
and C burning phases of stellar evolution where neutrons are 
produced by ($\alpha, n$) reactions on $^{13}$C and $^{22}$Ne.
While the $s$ process starts by neutron captures on iron seed
nuclei, neutron captures on the light isotopes 
present in the burning zones can initiate a neutron induced 
CNO cycle. Starting from the abundant $^{12}$C, the cycle is
represented by the neutron capture series on  
$^{12}$C, $^{13}$C, and $^{14}$C followed by the sequence
$^{15}$C($\beta^-$)$^{15}$N($n, \gamma$)$^{16}$N($\beta^-$)$^{16}$O($n, 
\gamma$)$^{17}$O($n, \alpha$)$^{14}$C or by producing $^{16}$N 
via
$^{14}$N($n, p$)$^{14}$C($n, \gamma$)$^{15}$C($\beta^-$)$^{15}$N($n, 
\gamma$).
The slowest reaction in this cycle is $^{14}$C($n, \gamma$)$^{15}$C
and, therefore, $^{14}$C can build up a correspondingly high
abundance. Although the outbreak from the neutron induced
CNO cycle via $^{17}$O($n, \gamma$)$^{18}$O is strongly 
suppressed by the dominance of the ($n, \alpha$) channel,
there might be a non-negligible effect on the neutron balance 
of the $s$ process depending on the cross section for 
$^{14}$C($n, \gamma$). To settle this issue, the cross section would
be needed for thermal energies in the 10 to 100 keV region. 

Neutron capture reactions on very light nuclei carry a substantial
part of the reaction flow in neutrino driven wind scenarios for the 
$r$ process \cite{TSK01}. Among these reactions $^{14}$C($n, 
\gamma$)$^{15}$C contributes mostly during the early phases, 
when $^{14}$C is formed via the sequences 
$^9$Be($\alpha, n$)$^{12}$C($n, \gamma$)$^{13}$C($n, \gamma$)$^{14}$C 
and 
$^9$Be($n, \gamma$)$^{10}$Be($\alpha, \gamma$)$^{14}$C. In these 
applications the ($n, \gamma$) cross section is required up to MeV 
energies because of the high temperatures in excess of $3\times 10^9$ 
K, corresponding to thermal energies of about 300 keV.

The $^{14}$C($n, \gamma$) reaction is also important to validate
the ($n, \gamma$) cross sections obtained by theoretical calculations \cite{WGT90,TBD06}
via the Coulomb dissociation
method in the experiments reported in Refs. \cite{HWG02,DAB03,NFA03}. 
In this approach the time-reversed 
process is measured via breakup of $^{15}$C projectiles in the 
virtual photon field of a $^{208}$Pb target. The ($n, \gamma$) cross 
section can then be inferred via detailed balance. The $^{14}$C($n, 
\gamma$)$^{15}$C reaction is the only case with neutrons so far where both,
direct and indirect,
approaches were investigated experimentally. In fact, $^{14}$C 
belongs to the few cases where the Coulomb dissociation method can be validated 
in a convincingly clean way.

The first direct measurement \cite{BWK92a} was carried out at
Forschungszentrum Karlsruhe using the same sample as in the present
experiment. At that time, however, the measurement was severely 
hampered by the fact that the nickel container used for the 
$^{14}$C powder sample had been strongly activated by a previous 
irradiation with 800~MeV protons. The present study is a repetition 
of this first measurement after a 12-year cooling time of the nickel 
container, which led to a reduction of this disturbing activity to 
an acceptable level. Another reason for repeating the 
measurement was that meanwhile a more efficient detector system for 
the induced activity became available. Finally, the energy range was significantly
extended compared to the previous experiment.

Preliminary results of this second experiment \cite{RHP05} turned out
to be subject to significant corrections resulting from the accidental
activation of the HPGe detector used. In this article we present a 
detailed description of the measurement as well as a thorough re-analysis 
of the data and of the remaining uncertainties. The results are compared
to model predictions and can be used to test the applicability of the
Coulomb dissociation method.

\section{Experiment}

\subsection{$\gamma$-detection}

The short half-life of $^{15}$C of only $t_{1/2}$=$2.449\pm0.005$~s \cite{AlM79,ABB97} necessitates 
the use of the
fast cyclic activation technique \cite{Bee91}. The induced activity
during each cycle was detected via the characteristic 5.2978~MeV $\gamma$-line
(relative intensity $I_\gamma~=~(63.2~\pm~0.8)\%$) in the $^{15}$C decay using a HPGe
detector with a relative efficiency of 100\%. 

The detector efficiency
was determined with a set of calibration sources and via the
$^{27}$Al(p,$\gamma$)$^{28}$Si reaction as described in
Ref.~\cite{AKH77}. The distance between the $^{15}$C sample and the HPGe detector
during the experiment was only 6~mm.  The distance between the detector
and the aluminum target during the $^{27}$Al(p,$\gamma$) calibration had
to be significantly increased to 76~cm in order to keep the probability
for summing of different $\gamma$-rays out of one cascade at a
negligible level. A thin layer of 148~nm aluminum, which corresponds to
40~$\mu$g/cm$^2$ as used by Anttila $et~al.$ \cite{AKH77}, was evaporated on a copper
backing. The thickness of the Al layer was determined with a quartz
crystal. The energy loss for 1-MeV protons in this layer is less than
7~keV.  The HPGe detector was placed at an angle of 55$^\circ$ as
described in Ref.~\cite{AKH77}.  

\begin{figure}
\includegraphics[width=20pc]{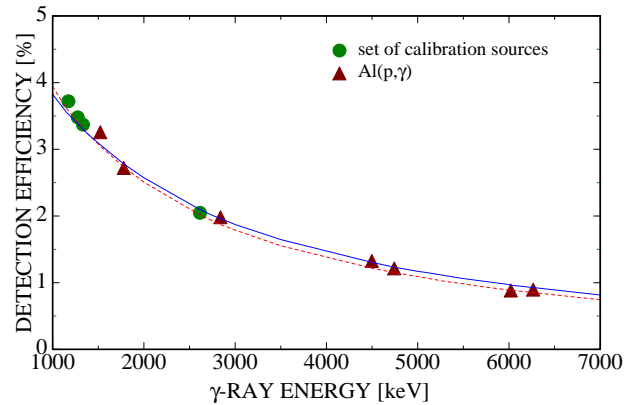}
\caption{(Color online) The $\gamma$-ray efficiency of the HPGe detector used during the cyclic activation. 
The efficiency was measured using calibration sources and the $^{27}$Al(p,$\gamma$)$^{28}$Si reaction. 
The red dashed and blue solid curves correspond to GEANT simulations
assuming 50~cm and 6~mm distance between sample and detector, respectively. 
The extrapolation from low to high energies results in slightly
smaller values for the 6~mm case.
\label{fig_g_eff}}
\end{figure}

The calibration at 76~cm was
completed with a set of calibrated sources, namely $^{22}$Na, $^{54}$Mn,
$^{60}$Co, and $^{208}$Tl (which is part of the $^{232}$Th decay
chain). A separate series of calibration measurements at a distance of 6~mm to the
detector was performed to normalize the above determined efficiency
curve to the geometry during the cyclic activation.  Very weak
calibrated samples of $^{54}$Mn, $^{65}$Zn, $^{88}$Y were used for that
purpose.  Simulations of the $\gamma$-ray efficiency using the detector simulation tool
GEANT~3.21 \cite{GEA93} showed
that the energy dependence is slightly different for the setup
during the activation (6~mm distance) and during the
$^{27}$Al(p,$\gamma$)$^{28}$Si experiment (50~cm), see
Fig.~\ref{fig_g_eff}. This effect was taken into account during the analysis.


The main difference to our preliminary analysis \cite{RHP05} was the
discovery of a huge dead time effect during the experiment. Even though
the detector was shielded from the neutron production target to reduce
radiation damage, enough neutrons reached the detector to produce
significant intrinsic $\gamma$ activity from ($n,\gamma$) 
reactions on $^{74}$Ge and $^{76}$Ge. The previous activation of the Ni
container led to the production of $^{44}$Ti in the container 
($t_{1/2}=60.3~\pm~1.3$~yr \cite{GMS98}). The 1.16~MeV $\gamma$-ray 
activity from the decay of the daughter $^{44}$Sc was 
measured independently and could be used as an
internal standard to determine the
crucial dead time corrections of about a factor of 3.

The decay properties of the radio-nuclides used in the analysis of the 
efficiency calibrations are summarized in Table~\ref{table:decay_properties}. 
The only important change compared to the previous experiment \cite{BWK92a} 
concerns the intensity of the 5297.8~MeV line in the decay of $^{15}$C. 
The new recommended value is (63.2~$\pm$~0.8)\% \cite{Fir96} instead of 
(68~$\pm$~2)\% \cite{BrF86}.

\begin{table}[htb]
\caption{Decay properties used for the determination of the detection
efficiency of the HPGe detector$^a$ and for evaluating the $^{15}$C activity \cite{Fir96}.}
\label{table:decay_properties}
\renewcommand{\tabcolsep}{2pc} 
\renewcommand{\arraystretch}{1.5} 
\begin{ruledtabular}
\begin{tabular}{@{}ccc}
 Isotope    & Energy       & Intensity \\
            & (keV)        & (\%)      \\
\hline
$^{15}$C    & 5297.8       & 63.2~$\pm$~0.8\\
$^{22}$Na   & 511          & 181.1     \\
            & 1274.5       &  99.94    \\
$^{44}$Sc   & 1157.0       & 99.9      \\
$^{54}$Mn   & 834.83       & 99.98     \\
$^{60}$Co   & 1173.2       & 99.9      \\
            & 1332.5       & 99.98     \\
$^{65}$Zn   & 1115.5       & 50.6      \\
$^{88}$Y    & 898.04       & 93.7      \\
            & 1836.0       & 99.2      \\
$^{208}$Tl  & 510.77       & 22.6      \\
            & 583.19       & 84.5      \\
            & 860.56       & 12.4      \\
            & 2614.5       & 99.16     \\
\end{tabular}
\end{ruledtabular}
$^a$The decay intensities of the calibration sources are known to better than ~0.5\% \cite{Fir96}.
\end{table}

A 1~mm thick lead sheet was placed in front of the HPGe detector during
the experiment as well as during the different calibrations in order to
reduce the strong low energy background caused by bremsstrahlung from
the $^{14}$C decay electrons.

The results of the efficiency calibration are summarized in
Table~\ref{table:g_eff}.  The photo-peak efficiency for the 5.2978~MeV line following the decay 
of \nuc{15}C was determined to (1.09~$\pm$~0.05)\%.

\begin{table}[htb]
\caption{Detection efficiencies for the 5.2978~MeV decay line of $^{15}$C.}
\label{table:g_eff}
\renewcommand{\arraystretch}{1.2} 
\begin{ruledtabular}
\begin{tabular}{@{}ccc}
 Process        & E (keV) & Efficiency (\%) \\
\hline
Double Escape (DE)  & 4275.8 & 0.18~$\pm$~0.04   \\
Single Escape (SE)  & 4786.8 & 0.58~$\pm$~0.06   \\
Full Energy (FE)    & 5297.8 & 1.09~$\pm$~0.05   \\
\hline
Sum of all above    & 		 & 1.86~$\pm$~0.09   \\
\end{tabular}
\end{ruledtabular}
\end{table}

\subsection{Neutron spectra}

Neutrons were produced via the $^7$Li($p,n$)$^7$Be reaction by bombarding 
metallic $^7$Li targets with proton beams provided by the Karlsruhe 3.7~MV
Van de Graaff accelerator. Different neutron energy distributions were 
obtained by varying of the proton energy and the thickness of the 
Li targets. 

The thickness of the \nuc{14}C sample in the neutron beam direction
was 5~mm . The neutron flux up- and downstream 
of the sample was monitored with two gold foils, allowing a measurement 
relative to the well known $^{197}$Au$(n,\gamma)$$^{198}$Au
cross section. At the end of each run the activity of the gold foils was
determined via the 412~keV $\gamma$-ray from the $^{198}$Au decay
($t_{1/2}$~=~2.7~d) using a well calibrated germanium detector.
The shape of the gold foils was 21x12~mm$^2$ according to the
activity distribution of \nuc{14}C in the sample, which was 
measured by detecting the emitted X-rays with a slit collimator.

\begin{figure}
\includegraphics[width=20pc]{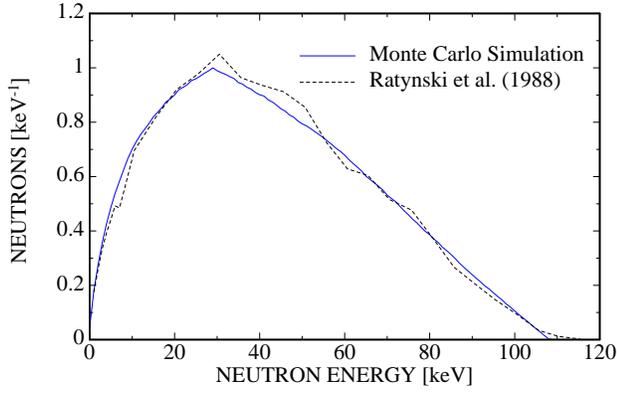}
\caption{(Color online) Comparison of a simulated neutron spectrum (solid, blue line) 
with experimental data (dashed, black line) \cite{RaK88}
in arbitrary units. }
\label{fig_25MACS_standard}
\end{figure}

The neutron fluxes obtained with the gold foils up- and downstream of 
the sample were significantly different due to the close geometry between 
the neutron source and the sample. This effect was evaluated by means of Monte-Carlo
simulations of the neutron spectra, starting from the double-differential 
\nuc{7}Li($p,n$) cross section from Liskien and Paulsen \cite{LiP75}
and including the energy loss of the protons in the lithium layer.
With this approach, the standard neutron spectrum used for activations 
\cite{RaK88} could be nicely 
reproduced as shown in Fig.~\ref{fig_25MACS_standard}.

\begin{figure}
\includegraphics[width=20pc]{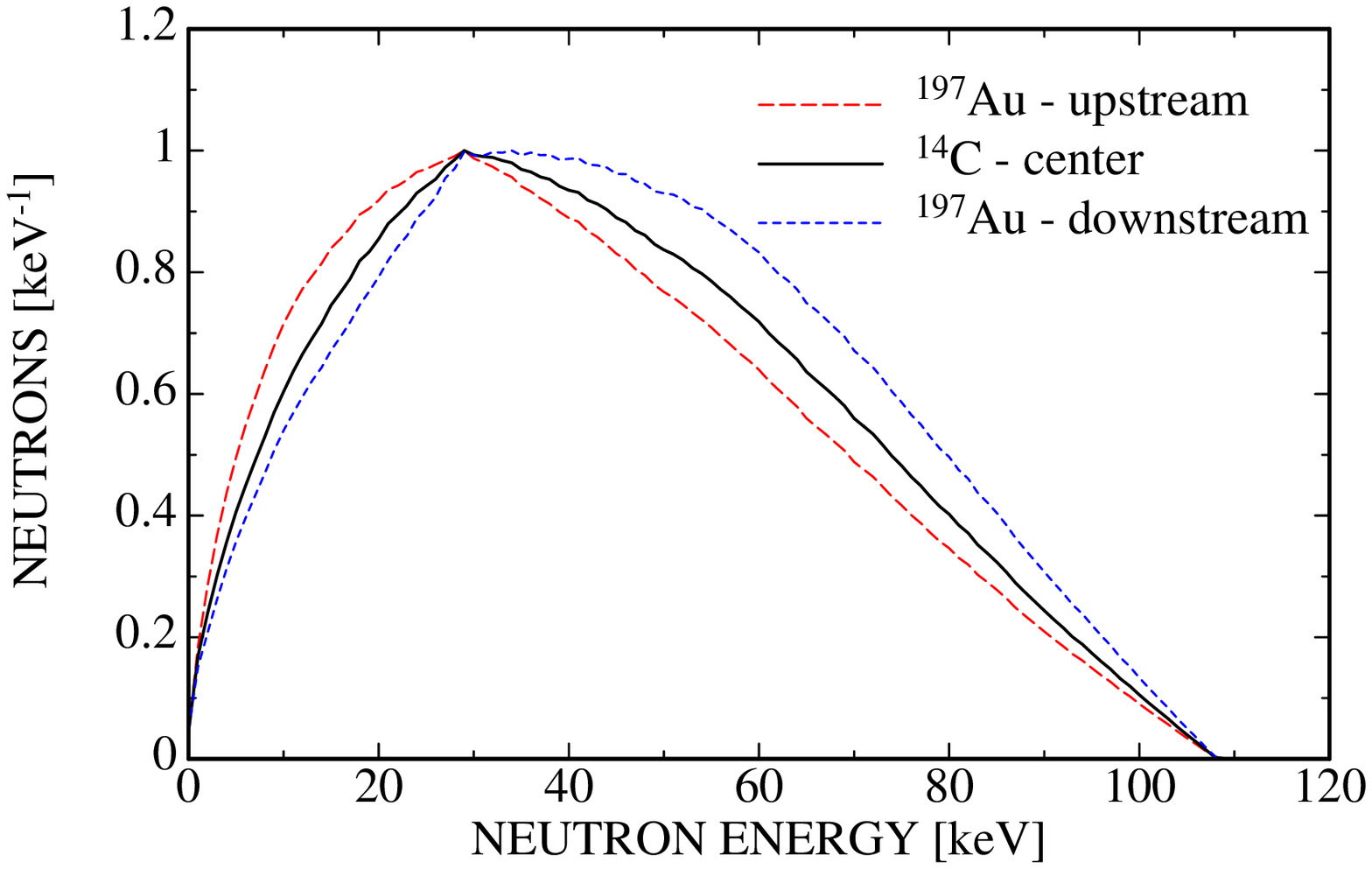}
\includegraphics[width=20pc]{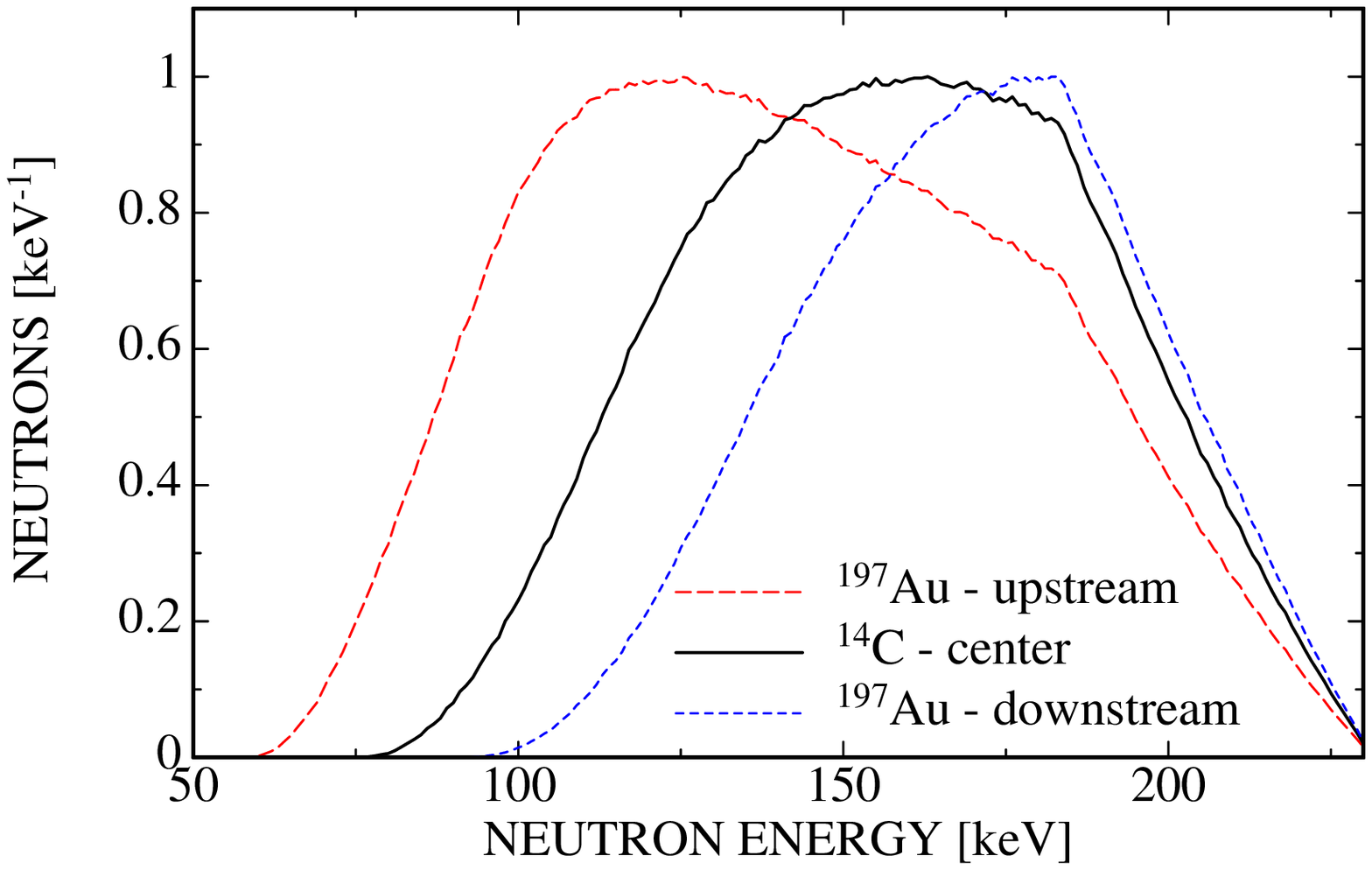}
\includegraphics[width=20pc]{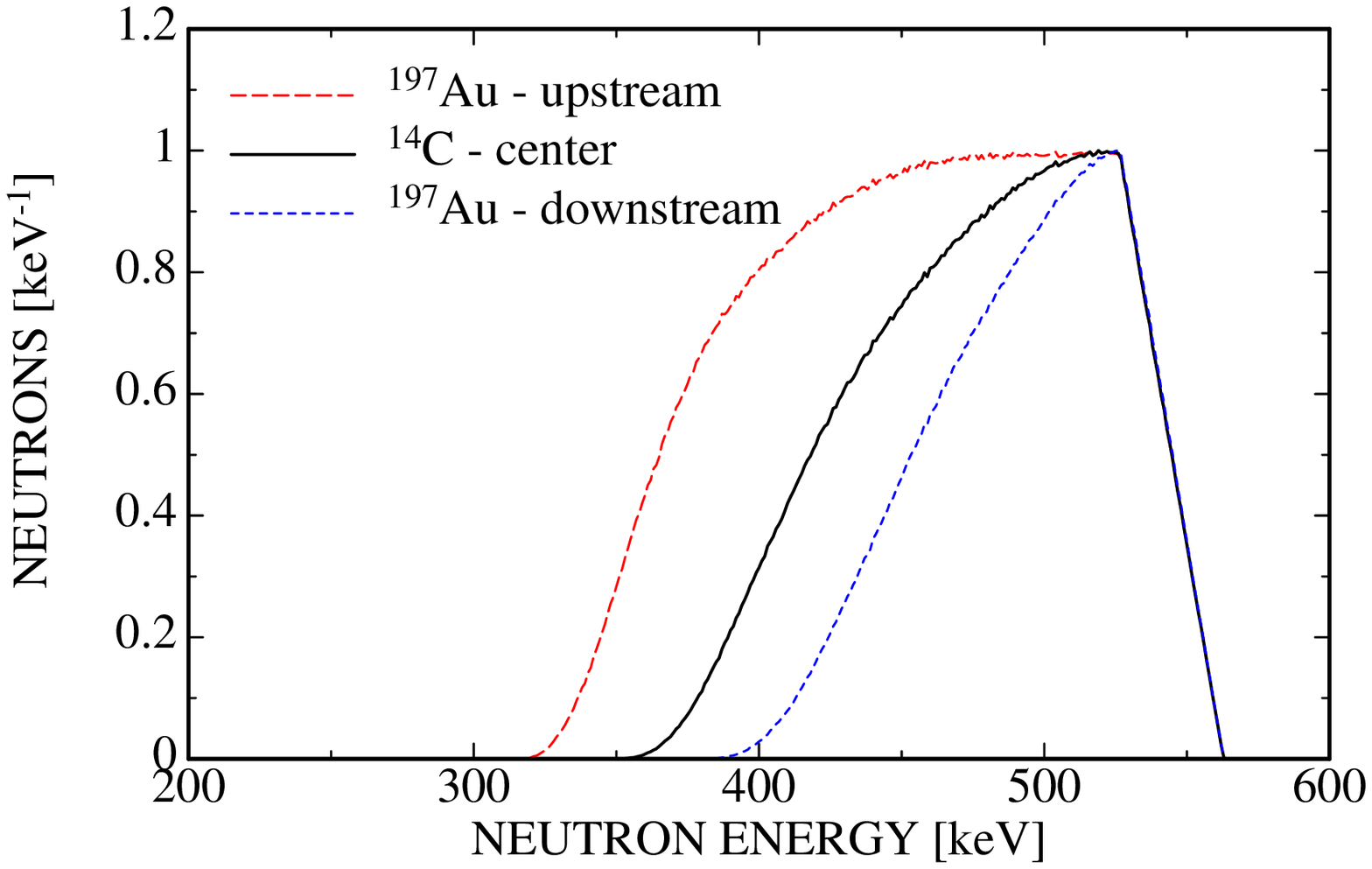}
\includegraphics[width=20pc]{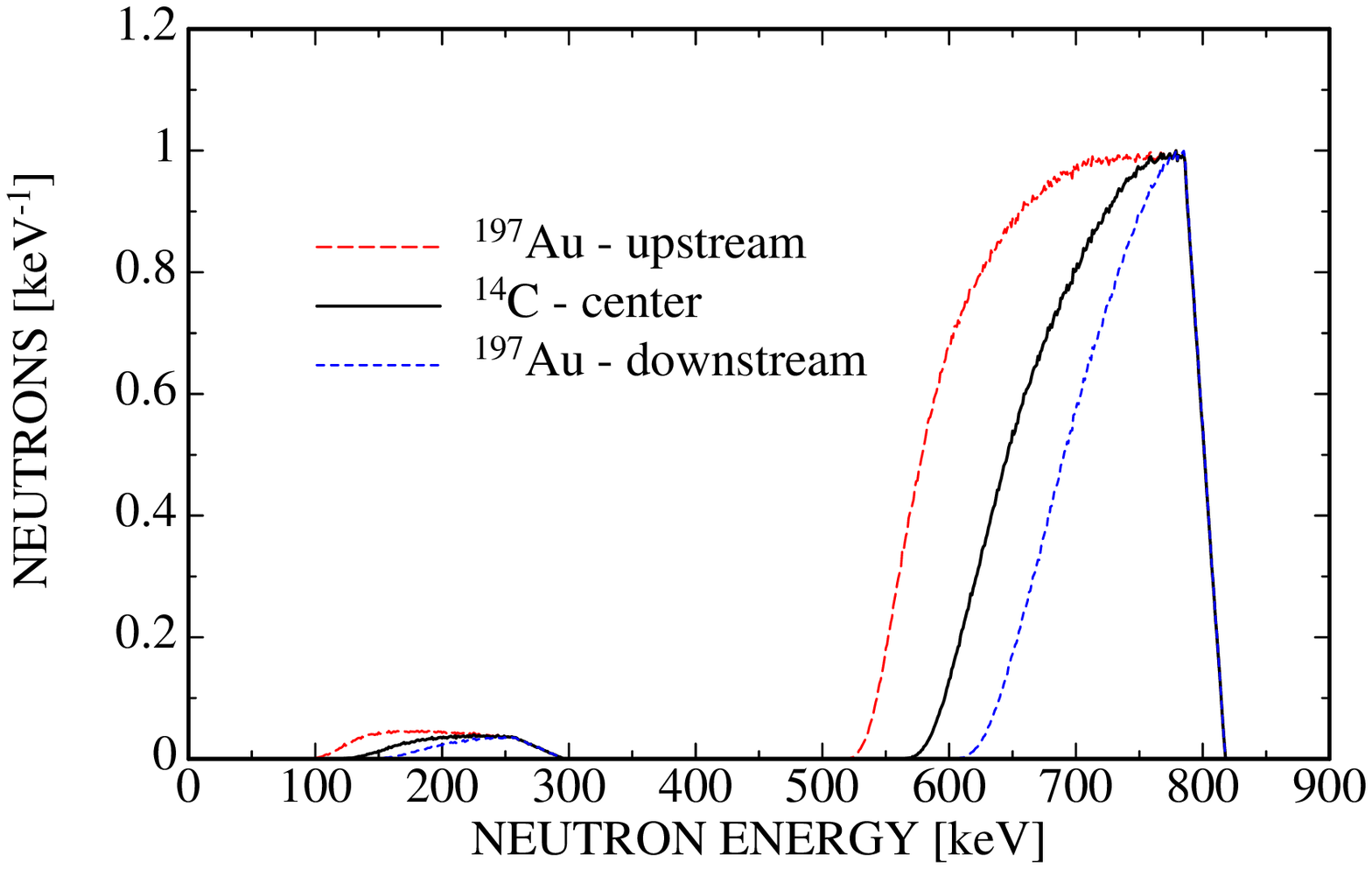}
\caption{(Color online) Neutron spectra for Runs~I-IV (top to bottom) in arbitrary units. The curves correspond to the
simulated spectra for the gold foils as well as for the \nuc{14}{C} sample. The second neutron group 
around 200~keV in the bottom panel results from the \nuc{7}{Li}(p,n)\nuc{7}{Be}${^*}$ reaction.}
\label{fig_neutron_spectra}
\end{figure}

Based on the good agreement for this spectrum, which is rather sensitive
to the proton energy distribution close to the neutron production
threshold, the same method was also applied to the runs at higher
energies. The resulting neutron energy distributions shown in 
Fig.~\ref{fig_neutron_spectra} represent effective spectra,
where the variation of the sample thickness as a function of 
neutron emission angle was properly considered. Correspondingly, 
the neutron spectra seen by the \nuc{14}C sample and by the gold foils
are exhibiting different widths. In addition, the spectra at 750~keV 
(bottom panel of Fig.~\ref{fig_neutron_spectra}) show a second neutron group around 200~keV, 
which results from the population of the first excited state in \nuc{7}Be 
at 429~keV \cite{LiP75}. The parameters of the different runs 
are summarized in Table~\ref{table:run_parameters}.
 
The corresponding integrated neutron fluxes are listed in 
Table~\ref{table:neutron_flux}. The values as derived from the gold foils 
were interpolated to the center plane of the sample in the
following way. Since the neutron spectra for the up- and downstream
gold samples differ significantly, an effective \nuc{197}{Au}$(n,\gamma)$
cross section had to be calculated in a first step. This was performed by folding
the effective neutron spectra with the differential \nuc{197}{Au}$(n,\gamma)$ cross
section by Macklin \cite{Mac82a}, normalized to Ratynski and K{\"a}ppeler \cite{RaK88},
during the Monte-Carlo simulations of the neutron spectra.
The neutron flux passing the up- and downstream gold samples were then determined
based on the number of produced \nuc{198}{Au} nuclei

\begin{equation*} 
\Phi = \frac{N_{198}}{N_{197}\cdot \sigma_{n,\gamma} \cdot f_b}
\end{equation*}

where the correction $f_b$ accounts for the fraction of $^{198}$Au
nuclei that decayed already during the irradiation \cite{BeK80}. 
The flux at the sample position and the corresponding systematic uncertainties
were then derived by normalizing the results from the Monte-Carlo simulations
to the measured neutron fluxes at the position of the gold foils:
\begin{eqnarray*}
  R_{up}		&=& {\Phi_{up}}/{n^{sim}_{up}} \\
  R_{down}		&=& {\Phi_{down}}/{n^{sim}_{down}} \\
  R				&=& \left({R_{up}+R_{down}}\right) /{2}  \\
  \frac{\mbox{d}R}{R} &=& \frac{R_{up}-R_{down}}{R_{up}+R_{down}} /2	\\
  \Phi_{sample} &=& R \cdot n^{sim}_{sample} 				\\
  \frac{\mbox{d}\Phi}{\Phi} &=&\frac{\mbox{d}R}{R} 
\end{eqnarray*}

\begin{table}[htb]
\caption{Beam parameters and activation times of the different runs.}
\renewcommand{\arraystretch}{1.2} 
\label{table:run_parameters}
\begin{ruledtabular}
\begin{tabular}{@{}ccccc}
Run             & $E_p$        & $d_{Li}$   & $E_n$  & $t_A$      \\
                & (keV)        & ($\mu{m}$) & (keV)  & (h)        \\
\hline
I               & 1912         & 30         & 23.3 (MACS) & 22.0  \\
II              & 2001         & 5          & 150 (average)  & 24.0  \\
III             & 2291         & 5          & 500 (average)  & 20.0  \\
IV              & 2530         & 5          & 750 (average)  & 5.5   \\
\end{tabular}
\end{ruledtabular}
\end{table}

\begin{table}[htb]
\renewcommand{\arraystretch}{1.5} 
\caption{Total neutron fluxes derived from the gold foils and interpolated 
to the center plane of the sample.}
\label{table:neutron_flux}
\begin{ruledtabular}
\begin{tabular}{@{}crlrlrl}
 Run & \multicolumn{2}{c}{Upstream}    & \multicolumn{2}{c}{Downstream}     & \multicolumn{2}{c}{Sample} \\
     & $\sigma{_{up}}$ & $\Phi{_{up}}$ & $\sigma{_{down}}$& $\Phi{_{down}}$ & $\Phi{_{sample}}$ & Uncert.\\
     & (mbarn)         & (10$^{13}$)   & (mbarn) & (10$^{13}$)              & (10$^{13}$)       & (\%)    \\
\hline
I    & 604  & 1.82                     & 546	& 1.26                      & 1.56           & 3.0\\
II   & 284  & 2.08                     & 270	& 0.62                      & 0.84           & 9.4\\
III  & 155  & 2.42                     & 133	& 1.09                      & 1.64           & 6.8\\
IV   & 106  & 0.95                     & 100	& 0.44                      & 0.655           & 6.0\\
\end{tabular}
\end{ruledtabular}
\end{table}

\subsection{Sample mass}
The independent determination of the sample mass by a calorimetric 
measurement of the decay heat turned out to be crucial for the 
analysis of this experiment. The use of this technique was favored 
by the comparably low $\beta^-$ end point energy of $E_{max}~=~156$~keV 
and by the fact that $^{14}$C decays without emission of $\gamma$-rays. 
The measurement was carried out at the Tritium Laboratory of 
Forschungszentrum Karlsruhe \cite{DBG05}, yielding a heat production 
of $370~\pm~4~\mu$W. Adopting an average energy $E_{avg}~=~49.475$~keV 
\cite{NUD07} for the decay electrons
and a half-life of $t_{1/2}~=~5700~\pm~30$~yr \cite{Fir96}, the measured 
decay heat corresponds to an activity of $1.26~\pm~0.01~$Ci or 
a total mass of $283~\pm~3$~mg of $^{14}$C. This value is 
independent of the isotopic enrichment (which was quoted to be 89\%)
and more than a factor of two less than the specified value,
which had been wrongly adopted in the previous activation \cite{BWK92a}.
This mismatch was presumably due to the undocumented removal of 
$^{14}$C powder from the original sample.

As mentioned before, the nickel container was still slightly active due
to its previous exposure to proton beams up to 800~MeV in energy. 
In principle, the measured decay heat of the sample represents, 
therefore, only an upper limit of the sample mass, since other 
radio-isotopes can contribute as well. However, a careful analysis 
of all potential candidates confirms that this correction can 
be neglected. The two main constraints would be the half-life 
and the $\gamma$-activity of the contaminating isotope. Since
the proton experiments were made 25 years ago, the half-life 
had to be in the range between 10 and 100~years, otherwise the 
isotope would have either already decayed or its specific activity 
would be too low to make any impact. The $\gamma$-activity was 
carefully measured by means of a HPGe detector with Be-window,
and was shown not to exceed the completely negligible level of the weak
$^{44}$Ti decay.

\subsection{Cyclic activation}

Each cycle consisted of an activation time of $t_{beam}$~=~10~s, the
$\gamma$-ray detection time $t_{det}$~=~10~s (during which the
proton beam was switched off), and twice the time for moving the sample
between detector and neutron production target $t_{wait}$~=~0.8~s.  

\begin{figure}
\includegraphics[width=20pc]{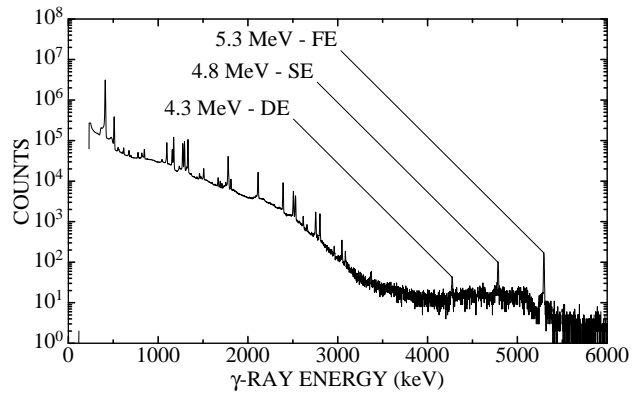}
\caption{Measured $\gamma$-ray spectrum after 22~h of cyclic activation
during Run I (23.3~keV MACS).
\label{fig_g_energy}}
\end{figure}

\begin{figure}
\includegraphics[width=20pc]{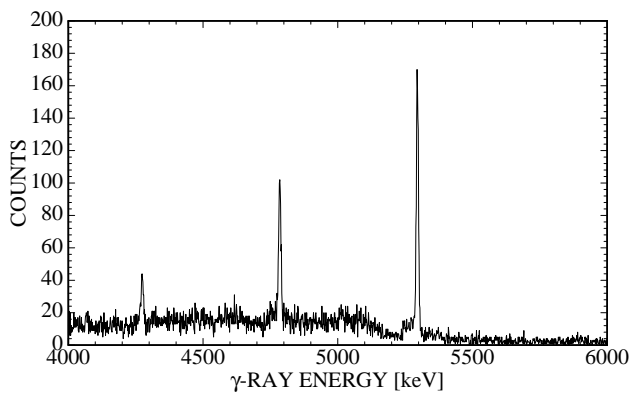}
\caption{The same data as shown in Fig.~\ref{fig_g_energy}, 
but focused on the decay lines of $^{15}$C.
\label{fig_g_energy_zoom}}
\end{figure}

\begin{figure}
\includegraphics[width=20pc]{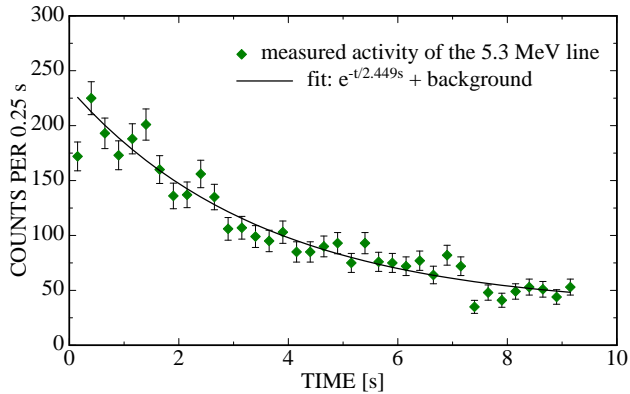}
\caption{(Color online) Time dependence of the $^{15}$C activity.
\label{fig_g_tof}}
\end{figure}

Figs.~\ref{fig_g_energy} and~\ref{fig_g_energy_zoom} show a typical 
$\gamma$-ray spectrum taken during the experiment. Full energy, single 
escape as well as double escape peaks of the 5.2978~MeV line from the 
decay of $^{15}$C are obtained with good signal/background ratios. 
In order to reduce systematic uncertainties, the time
dependence of the $^{15}$C decay during the 10~s counting period, has
been monitored. The decay curve is compared in Fig.~\ref{fig_g_tof} with
a fit assuming a constant background and the exponential decay law with
$2.449\pm0.005$~s half-life. Within the statistical uncertainties the 
measured activity follows the expected time dependence.

\section{Analysis and Results}
\subsection{Results from the activation measurement}

The \nuc{14}{C}$(n,\gamma)$ cross section for the different neutron spectra 
were determined from the ratio of \nuc{15}{C} to \nuc{14}{C} atoms including corrections
for \nuc{15}{C} atoms decayed, while the sample was not in front of the HPGe detector. The 
details of this method are very well described in \cite{BRW94,BWK92a}.
The results including uncertainties are presented in Table~\ref{table_results}.
The main contributions to the overall uncertainty come from
counting statistics (2-8\%), the $\gamma$-ray detection efficiency
(5\%), and the determination of the neutron flux (2-10\%). All other
uncertainties are smaller than 2\%. With respect to the second neutron group during 
Run~IV (750~keV), we performed the entire analysis three times. A first time without consideration
of the second group, a second time with the cross section as suggested by \cite{LiP75},
and a last time adopting twice the cross section populating the excited state of \nuc{7}{Be}.
The value for the resulting \nuc{14}{C}$(n,\gamma)$ cross section increased by 5\% in each step. Therefore
we quadratically added an additional systematic uncertainty of 5\% to the cross section at 750~keV.


\begin{table*}[htb]
\caption{Cross sections of the $^{14}$C$(n,\gamma)$$^{15}$C 
reaction.  }
\label{table_results}
\renewcommand{\arraystretch}{1.2} 
\begin{center}
\begin{ruledtabular}
\begin{tabular}{@{}ccccc}
Run   & Neutron energy     &  \multicolumn{3}{c}{Cross section results ($\mu$barn)}\\
\cline{3-5}
      & distribution (keV) &  Measured values$^a$  & Theory$^b$     & Theory/Experiment \\
\hline
I     & 23.3 (MACS)        & 7.1$\pm$0.5 (6.7)     & $6.5 \pm 0.4$  & $0.92 \pm 0.08$   \\
II    & 150  (average)              & 10.7$\pm$1.2 (11)     & $11.7 \pm 0.6$ & $1.09 \pm 0.12$   \\
III   & 500  (average)              & 17.0$\pm$1.5 (8.8)    & $16.5 \pm 0.8$ & $0.97 \pm 0.10$   \\
IV    & 750  (average)              & 15.8$\pm$1.6 (10)     & $17.5 \pm 0.9$ & $1.11 \pm 0.11$   \\ 
\end{tabular}
\end{ruledtabular}
\end{center}
$^a$ Relative uncertainties ( in \%) are indicated in brackets.\\
$^b$ Convoluted with the neutron spectra of Fig.~\ref{fig_neutron_spectra}.
\end{table*}

\subsection{Theoretical modeling of direct radiative capture%
\label{sec:theory}}
We used a simple potential model to calculate the cross section for
direct capture of a low-energy neutron on \nuc{14}{C}. The modeling of
this radiative process was further simplified by relying on Siegert's
theorem to approximate the exact current form $(\mathbf{j} \cdot
\mathbf{A})$ of the electromagnetic operator by its density form in
terms of electrostatic multipoles. The calculations were performed with
the direct-reaction code FRESCO~\cite{Tho88}, and the radiative capture
was modeled as a one-step process using first-order DWBA theory. A real
potential was used to describe the incident wave, which is appropriate
for capture far from resonances, and to generate the single-particle
configurations of the ground and first-excited states of \nuc{15}{C}
(bound $1s_{1/2}$ and $0d_{5/2}$, respectively). These two states are
very close to pure single-particle configurations, which validates our
potential-model approach. Since we assume that the observed,
non-resonant cross section corresponds to direct radiative capture, the
calculated cross section only had to be normalized by the final
bound-state spectroscopic factor. The ratio of experimentally observed
to calculated cross section is then a measure of the spectroscopic
purity of the single-particle configuration. We
note that the particular structure of the \nuc{15}{C} states implies
that E1 capture is only possible for p-wave neutrons. The possibility of
E2 capture of s-wave neutrons to the first-excited state was also
included in these calculations, but the contribution to the capture
cross section was found to be less than $5\%$ at the relevant energies.

In all calculations, single-particle configurations were generated from
a Woods-Saxon potential well with the geometry of Ref.~\cite{CBM03}. The
potential depths were chosen to reproduce the binding energies of the
two bound states in \nuc{15}{C} with respect to the $\nuc{14}{C} + n$
thresholds. This procedure led to slightly different potential depths
for the $l=0$ ($V_{l=0} = 52.81$~MeV) and $l>0$ ($V_{l>0} = 51.33$~MeV)
channels. Since $p$-wave capture is the dominating process, the $l>0$
potential was used to describe the scattering wave of the incoming
channel. The use of $l$-dependent potentials is, in principle, not
compatible with the requirements of applying Siegert's theorem. However,
for the case considered here, we found that the difference between the
initial- and final-state potentials was so small that Siegert's theorem
was still valid.

\begin{figure}
\begin{center}
\includegraphics[width=20pc]{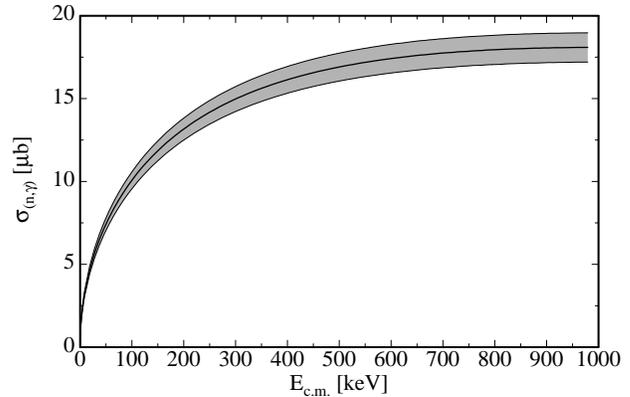}
\end{center}
\caption{Theoretical \nuc{14}{C}$(n,\gamma)$\nuc{15}{C} cross section,
  fitted to the experimental data, as described in the text.
\label{fig_results_theory}}
\end{figure}

The calculated radiative-capture cross section was convoluted with the
neutron spectra of Fig.~\ref{fig_neutron_spectra} 
to facilitate a direct comparison with the data from
the activation measurement.  The calculated capture to the first excited
state of \nuc{15}{C} was normalized by the spectroscopic factor $C^2S_1
= 0.69$, extracted from experimental neutron transfer $\nuc{14}{C}(d,p)\nuc{15}{C}^*$
data~\cite{GJB75}. Since this channel contributes less than
$5\%$ to the total capture cross section at the relevant energies, the
final result is not very sensitive to the particular choice of this
spectroscopic factor.  A fit to the experimental data, weighted by the
relative error bar of each data point, was then performed and resulted
in a best-fit spectroscopic factor of $C^2S_0 = 0.95 \pm 0.05$ for the
ground state $1s_{1/2}$ single-particle configuration, which is in good agreement
with 0.88 as derived from (d,p) data~\cite{GJB75}. The final
calculated cross section, convoluted with the different neutron spectra,
is compared with the experimental data in
Table~\ref{table_results}. In addition, the energy-differential
cross section, including the 1$\sigma$ error band, is shown in
Fig.~\ref{fig_results_theory}.\\
\subsection{Recommended astrophysical reaction rates}
\begin{figure}
\begin{center}
\includegraphics[width=20pc]{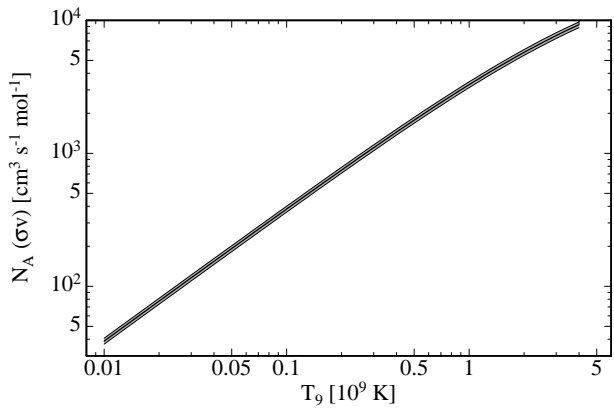}
\end{center}
\caption{Reaction rates for \nuc{14}{C}$(n,\gamma)$\nuc{15}{C}
  as a function of stellar temparature $T_9$ given in $10^9$~K.
\label{fig_results_rate}}
\end{figure}

The fitted theoretical cross section of the previous section was used to
compute reaction rates for astrophysical applications. The resulting
reaction rate is plotted in Fig.~\ref{fig_results_rate} as a function of
stellar temperature $T_9$ (in units of $10^9$~K). The applicability of the
calculated capture cross section is restricted by the experimental
energy range used in the activation measurement, i.e. from 1~keV to
1~MeV. The extracted reaction rate is therefore presented up to a
maximum temperature of $4 \cdot 10^9$~K. Extrapolations beyond this
temperature range would yield results that are not restricted by the
data from the present experiment.

The reaction rates were fitted to the parametrization suggested by
Rauscher and Thielemann~\cite{RaT00}
\begin{equation*}
  \begin{split}
  N_A \langle \sigma v \rangle =
  \exp \left(
  a_1+a_2 T_9^{-1}+a_3 T_9^{-1/3}+a_4 T_9^{1/3} \right.\\
  \left. +a_5 T_9+a_6 T_9^{5/3}+a_7 \ln(T_9)
  \right).
  \end{split}
\end{equation*}
The reaction rate is given in cm$^3$s$^{-1}$mol$^{-1}$ with the
temperature in $10^9$~K. The best-fit parameters, which  
reproduce the numerical values to within $0.01\%$ in the $0.01 \leq T_9
\leq 4.0$ temperature range, are:
\begin{eqnarray*}
  a_1 &=& 0.850\cdot 10^{1}		\\
  a_2 &=& -0.305\cdot 10^{-3} 	\\
  a_3 &=&  0.580\cdot 10^{-1} 	\\
  a_4 &=& -0.355\cdot 10^{0} 	\\
  a_5 &=& -0.116\cdot 10^{0} 	\\
  a_6 &=&  0.122\cdot 10^{-1} 	\\
  a_7 &=&  0.109\cdot 10^{1} 	\\
\end{eqnarray*}

\section{Discussion and Astrophysical Implications}

Compared to the result of the previous activation with $kT~=23.3~$keV
\cite{BWK92a} ($1.72~\pm~0.43~\mu$b) we find agreement, if the sample
mass measured in
this work and the currently available decay properties
of \nuc{15}{C} are taken into account. The agreement is then within 1$\sigma$.

\begin{figure}
\begin{center}
\includegraphics[width=20pc]{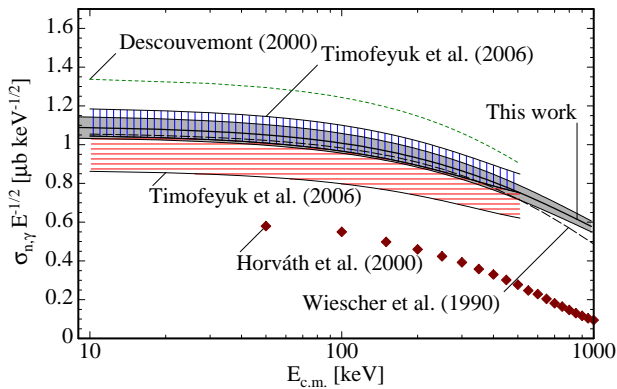}
\end{center}
\caption{(Color online) Comparison between the present results and
previous data.
\label{fig_results_differential}}
\end{figure}

All available differential data for the total capture cross section
of $^{14}$C are compared in Fig~\ref{fig_results_differential}. The data
are divided by $\sqrt{E}$ to remove the energy dependence caused by the
$p$-wave orbital-momentum barrier. The present cross section results
are in good agreement with theoretical estimates of Wiescher {\it
et al.} \cite{WGT90} and with the recently published estimates of
Timofeyuk {\it et al.} \cite{TBD06} based on mirror symmetry considerations.
Our data fall approximately 20\% below the values of Descouvemont
\cite{Des00}, but exhibit the same energy dependence.

The results of Horv{\'a}th {\it et al.} \cite{HWG02}, which were
obtained in a Coulomb-breakup study, show a large, constant offset
(Fig~\ref{fig_results_differential}). In other words, not only the
cross section values are different, but also the energy dependence.
The difference can be expressed as:
\begin{eqnarray*}
  \sigma_{present} &=& \sigma_{Horvath} + c \cdot \sqrt{E_{c.m.}}
         \\
 \mbox{with}~c &=&0.48~\mu\mbox{b/keV}^{1/2} \\
\end{eqnarray*}
With respect to the importance of the $^{14}$C($n,\gamma$)$^{15}$C
cross section for validating the Coulomb-break-up approach for
deducing this cross section from the time-reversed dissociation of
$^{15}$C it is important, however, to emphasize that the present
results are in good agreement with preliminary data from two other
Coulomb break-up studies \cite{DAB03,NFA03,Nak04}.

\begin{figure}
\begin{center}
\includegraphics[width=20pc]{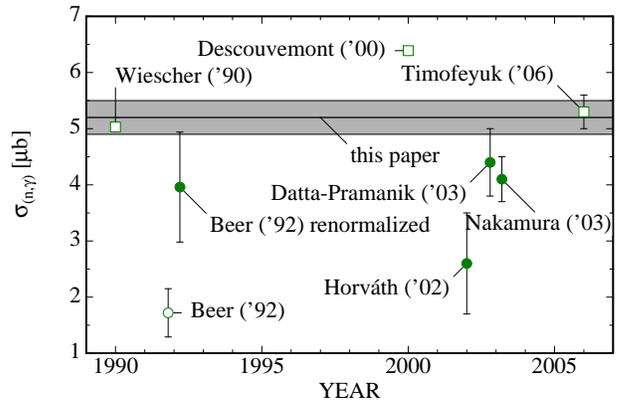}
\end{center}
\caption{(Color online) Comparison between this measurement (shaded
band) and previous
cross section results at $E_{c.m.}~=~23.3~$keV. Open squares refer to theoretical estimates 
while full circles refer to experiments including Coulomb-breakup studies. The only open
circle refers to the measurement by Beer {\it et al.} before the renormalization based on the new mass 
and line intensity information (see text). The respective references from left to right are 
\cite{WGT90},\cite{Des00},\cite{TBD06} (theoretical) and \cite{BWK92a},
\cite{HWG02},\cite{Pra02,DAB03},\cite{NFA03,Nak04} (experimental).
\label{fig_results_23kev}}
\end{figure}

Since the paper by Beer {\it et al.} \cite{BWK92a}, a comparison of the
differential
cross section at 23.3~keV is published in most papers dealing with the
\nuc{14}C($n,\gamma$) cross section. We note that the value published by
Beer {\it et al.} was a Maxwellian averaged cross section for
$kT~=~23.3~keV$,
which is different from the differential cross section at
$E_{c.m.}~=~23.3~$keV.
In  this tradition, a comparison of the differential 23.3~keV cross
sections is presented in Fig.~\ref{fig_results_23kev}. The present value
of 5.2~$\pm$~0.3~$\mu$barn is based on the theoretical description of the
cross section provided in the previous section.

The rate suggested by \cite{WGT90} has been used for most of the
nucleosynthesis simulations of the scenarios summarized at the
beginning of this paper. The agreement with the present
experimental results confirms many of the previous model
predictions. While present Cosmologies dismiss the likelihood of
inhomogeneous Big Bang scenarios, previous simulations of the
associated nucleosynthesis \cite{RAC94} based on this
$^{14}$C($n, \gamma$) reaction rate demonstrated a substantial
production of $^{14}$C at such conditions.

The role of the $^{14}$C($n, \gamma$)$^{15}$C reaction as the slowest 
link in the neutron induced CNO cycles proposed by \cite{WGS99} is also
confirmed by the present results. Detailed simulations now help to
analyze the impact of such a cycle on the neutron flux during core
carbon burning and shell carbon burning. These results indicate
that many more branches exist due to the presence of charged
particles in stellar helium and carbon burning environments
\cite{PGH07}. For helium burning most of the $^{13}$C produced by
$^{12}$C($n, \gamma$) is depleted by the $^{13}$C($\alpha, n$)
reaction rather than by $^{13}$C($n, \gamma$) and the production of
$^{14}$C is negligible as shown already by \cite{TAG96}. This may
be different for shell carbon burning which is characterized by
higher $^{12}$C abundances and a significantly lower $\alpha$
flux. New simulations on aspects of neutron production and capture
reactions are presently in preparation 
\cite{PGW08}. The study indicates that the main production of
$^{14}$C is given by the two reactions $^{14}$N($n, p$)$^{14}$C and
$^{17}$O($n, \alpha$)$^{14}$C. Because of the here confirmed low
cross section, the $^{14}$C($n,\gamma$) reaction does not play a
significant role for reducing the $^{14}$C abundance. However,
because of the relatively high temperatures of T$\approx$1 GK in the
carbon burning zone, alternative depletion channels open via 
$^{14}$C($p, n$)$^{14}$N with a negative Q-value of -626 keV and via 
$^{14}$C($\alpha,\gamma$)$^{18}$O alpha capture providing a new
abundance balance.

New simulations are also underway for studying the impact of
neutron capture reactions on neutron rich Be, B, and C isotopes on
the nucleosynthesis of light elements in neutrino driven wind
supernova shock scenarios \cite{BGM06}. The completion of these
studies does however require a detailed analysis of neutron
capture reactions on short-lived neutron rich isotopes to simulate
the anticipated reaction flow reliably \cite{TSK01}. New shell
model based simulations of these rates are presently in
preparation taking also into account the rapidly growing
experimental nuclear structure information on neutron rich nuclei
in the Be to Ne range.

\section{Summary}

We have measured the $^{14}$C($n,\gamma)$$^{15}$C cross section applying
the activation technique with four different energy distributions.
The results of the present experiment has removed the uncertainty
associated with the results of the previous $^{14}$C($n,\gamma$)
activation measurement by \cite{BeK80}. A theoretical fit of the
present cross section data is in good agreement with the capture
cross sections deduced from Coulomb dissociation studies of
$^{15}$C beams by \cite{DAB03} and \cite{NFA03} while in striking
disagreements with a third measurement by \cite{HWG02}. Moreover,
our results and analysis demonstrate good agreement with a number
of theoretical predictions for the reaction rate by \cite{WGT90}
and \cite{TBD06}.

The experimental results confirm the rate suggested by 
\cite{WGT90}, which has been used for most of the nucleosynthesis 
simulations mentioned in the Sec. I. Therefore, the astrophysical 
consequences of the previous model predictions remain essentially 
unchanged. A new aspect concerning the role of neutron capture 
reactions on neutron rich Be, B, and C isotopes is the production 
of light elements in neutrino driven wind supernova shock scenarios, 
which are presently under investigation \cite{BGM06}.

\begin{acknowledgments}
We would like to thank E.-P.~Knaetsch, D.~Roller, and W.~Seith for
their support at the Karlsruhe Van de Graaff accelerator. We are
also grateful to M.~Pignatari for discussing the impact of our
experimental results on the $^{14}$C nucleosynthesis during
stellar helium and carbon burning. This work was partly supported
by the Joint Institute for Nuclear Astrophysics (JINA) through NSF
Grants Nos. PHY-0072711 and PHY-0228206, and partly performed
under the auspices of the U.S. Department of Energy by the
University of California, Lawrence Livermore National Laboratory
(LLNL) under contract No. W-7405-Eng-48 and the Los Alamos 
National Laboratory (LANL) under the auspices of Los Alamos National
Security, LLC, DOE contract number DE-AC52-06NA25396.

\end{acknowledgments}

\newcommand{\noopsort}[1]{} \newcommand{\printfirst}[2]{#1}
  \newcommand{\singleletter}[1]{#1} \newcommand{\swithchargs}[2]{#2#1}

\end{document}